\documentclass[a4paper,11pt]{cambridge6A}
\pdfoutput=1 

\usepackage{amssymb}

\usepackage{bbm}

\usepackage[T1]{fontenc} 
\begin{document} 
\begin{titlepage}
	\thispagestyle{empty}
	
	\vspace{35pt}
	
	\begin{center} 
{ \LARGE{\bf No-scale supergravity}}		
		\vspace{40pt}
		
		{\LARGE Fabio~Zwirner}~\footnote{\large fabio.zwirner@pd.infn.it}
				
		\vspace{15pt}
		
			{{\it  Dipartimento di Fisica e Astronomia ``Galileo Galilei''\\
		Universit\`a di Padova, Via Marzolo 8, 35131 Padova, Italy}
		
		\vspace{10pt}

and

		\vspace{10pt}
		
	        {\it   INFN, Sezione di Padova \\
		Via Marzolo 8, 35131 Padova, Italy}
		
		}

		\vspace{40pt}
		
		{ABSTRACT}
	\end{center}

To connect supergravity with the real world, a highly non-trivial requirement is complete spontaneous supersymmetry breaking in an approximately flat four-dimensional space-time.
In no-scale supergravity models, this naturally happens at the classical level: the gravitino mass, setting the scale of supersymmetry breaking, slides along a flat direction of the potential with vanishing energy.
This contribution briefly describes, with a personal selection of simple illustrative examples, some qualitative features of no-scale models that relate them to a possible dynamical generation of the hierarchies between the vacuum energy scale, the weak scale and the Planck scale.
It includes comments on their versions with extended supersymmetry, on their higher-dimensional origin and on how their still unsolved problems of quantum stability can already be addressed, with some results, at the level of supergravity compactifications, although their solution (if any) will eventually require a better understanding of superstring theories. 

\vspace{30pt}

\begin{center}
Invited contribution to the book {\em Half a century of Supergravity},  \\  A.~Ceresole and G.~Dall'Agata eds., Cambridge University Press
\end{center}

\end{titlepage}

\chapter{No-scale supergravity} 
\label{ch:noscale}

\vspace*{-4.3cm}
\begin{center}
{\small Fabio Zwirner~\footnote{Dipartimento di Fisica e Astronomia ``Galileo Galilei'', Universit\`a di Padova and INFN, Sezione di Padova, 35131 Padova, Italy, fabio.zwirner@pd.infn.it}}
\end{center}
\vspace*{0.1cm}
\begin{center}
{\it Dedicated to Costas Kounnas, who introduced me to no-scale supergravity}
\end{center}
%
%
%
%
%
%

\section{Introduction} 
\label{sec:introduction}

In the physics of the fundamental interactions, there are two outstanding unsolved hierarchy problems: the smallness of the Fermi scale of weak interactions, $G_F^{-1/2} \sim 10^{-16} \, M_P$, and even more the smallness of the present vacuum energy density, $\rho_V \sim (10^{-30} \, M_P)^4$, with respect to the Planck scale of gravitational interactions, $M_P = (8 \pi G_N)^{-1/2} \simeq 2.4 \times 10^{18} \, {\rm GeV}$, as defined in terms of Newton's constant $G_N$.

For a long time it has been suspected that the beautiful mathematical structure and the special ultraviolet properties of supersymmetric theories may be eventually related to the solutions of the above two problems. 
In particular, at the end of the 20th century it was expected by many that the gauge hierarchy problem would be solved by the existence of supersymmetric partners of the Standard Model (SM) particles with masses of the order of the Fermi scale. 
This expectation has not been borne out by experiment: a light weakly coupled Higgs boson with mass $m_h \simeq 125 \, {\rm GeV}$ has been found at the LHC, with all its properties measured so far compatible with those predicted by the SM, and no confirmed direct or indirect signal of supersymmetric particles has been detected in a broad spectrum of experimental searches, whose sensitivity goes significantly beyond the Fermi scale.      

Still, space-time supersymmetry is an important ingredient of our best candidate for a consistent quantum theory of all fundamental interactions, including gravity: superstring theory. 
Perturbative formulations of superstring theory contain a single fundamental mass scale, the string scale $M_S$, and additional spatial dimensions.
The first step to make them realistic is to consider solutions where the extra dimensions are compact.
The Planck scale $M_P$ and the compactification scale(s) $M_C$ depend on the vacuum expectation values of certain fields present in the theory and should be determined by some dynamics on which we do not have full control at present. 
Supergravity theories are in general non-renormalizable (the issue is not completely settled for maximal supergravity, which could be finite), therefore they can be regarded as effective low-energy theories valid at energies smaller than $M_S$, if supersymmetry is not fully broken at the string scale, and also smaller than $M_C$, if supersymmetry is not fully broken in the compactification to $d=4$ space-time dimensions.        
    
In any realistic supergravity model, supersymmetry should be completely broken, as we do not observe supersymmetric partners of the known particles, and the breaking should be spontaneous, through the super-Higgs effect \cite{Volkov:1973jd, Deser:1977uq, Cremmer:1978hn}. 
Since the general scalar potential of supergravity is not positive definite, the question then arises of whether local supersymmetry can be completely broken with vanishing vacuum energy, i.e. on a flat Minkowski background.
In the first models of $N=1$, $d=4$ spontaneously broken supergravity \cite{Polonyi:1977pj}, this was achieved by a fine tuning of the model parameters.   

In 1983 Cremmer, Ferrara, Kounnas and Nanopoulos \cite{Cremmer:1983bf} (see also \cite{Chang:1983hk}) gave the first examples of $N=1$, $d=4$ supergravity models where, at the classical level: 
supersymmetry is broken with vanishing vacuum energy on a continuum of inequivalent degenerate vacua; 
the gravitino mass $m_{3/2}$, setting the supersymmetry-breaking scale in Minkowski space, slides along such flat direction in field space;
the potential is positive semi-definite, so that when moving along the flat directions there are no classical instabilities in other directions.
The term {\em no-scale models} appeared soon after, in some papers \cite{Ellis:1983sf,Ellis:1983ei,Ellis:1984bm} that  tried to explore how the hierarchy $m_W \sim m_{3/2} \ll M_P$ could be dynamically generated~\footnote{In those early explorations the contributions to the effective potential of order $m_{3/2}^4$ were omitted, and included only later \cite{Kounnas:1994fr}.}. 
Since then, such term has been used both in the supergravity literature and (often abusively) in the phenomenological literature on extensions of the SM with softly broken supersymmetry. 
In the following I will use it only for supergravity models with the three classical properties described above, or at least the first two.

{}From the early days of no-scale models, it was realised (see e.g. \cite{Polchinski:1985hi}) that they must address a number of stability issues.
If the $m_W/M_P$ hierarchy is to be generated by infrared logarithmic quantum corrections via dimensional transmutation \cite{Coleman:1973jx,Weinberg:1973ua}, then quantum corrections to the effective potential should be at most ${\cal O} (m_{3/2}^4)$, not  ${\cal O} (m_{3/2}^2 M_P^2)$ as expected on general grounds in spontaneously broken $N=1$ supergravity.
A much softer ultraviolet behaviour is required, to leave room for this possibility, in models where $m_{3/2} \gg m_W$, as well as when we want to address the approximate stability of the flat Minkowski background.
Despite some progress along these lines, to be described later, we are still far from a convincing solution.

With a personal selection, which will unavoidably neglect many important works, the rest of this contribution will be focused on some aspects of no-scale supergravity that are standing the test of more than four decades of theoretical and experimental progress and are still relevant for current research.



%
%
%
\section{Four-dimensional no-scale models}
\label{sec:4dns}

\subsection{Basics of $N=1$, $d=4$ supergravity}
\label{subsec:basics}

A $N=1$, $d=4$ supergravity model with chiral multiplets $\phi^i \sim (z^i, \psi^i)$ and vector multiplets $U^a \sim (\lambda^a, A_\mu^a)$ is specified by three ingredients~\footnote{Here and in the following, we use natural units where $M_P=1$.}. 
The first is the real and gauge-invariant function
\begin{equation}
G = K + \log |W|^2 \, ,
\label{ggen}
\end{equation}
where $K$ is the real K\"ahler potential and $W$ the holomorphic superpotential. 
If we are not interested in the gauging of the R-symmetry, leading to constant Fayet--Iliopoulos terms, we can assume that both $K$ and $W$ are gauge invariant.  
The second is the holomorphic gauge kinetic function $f_{ab}$. Generalized Chern--Simons terms may also be needed, 
but they will not play any r\^ole here. 
The third are the holomorphic Killing vectors $X_a = X_a^i (z) (\partial/\partial z^i)$, which generate the analytic isometries of the K\"ahler manifold for the scalar fields that are gauged by the vector fields. 
In the following, for simplicity, we will always denote $G$, $f_{ab}$ and $X_a$ as functions of the complex scalars $z^i$ rather than the superfields $\phi^i$.

The gauge transformation laws and covariant derivatives for the scalars in the chiral multiplets read
\begin{equation}
 \delta z^i  =  X_a^i \, \epsilon^a \, , 
\qquad
D_\mu z^i = \partial_\mu z^i - A^a_\mu X_a^i \, ,
\end{equation}
where $\epsilon^a$ are real parameters. 
The classical scalar potential is made of three contributions, controlled by the auxiliary fields of the gravitational, chiral and vector multiplets: 
$$
V_0 = V_G + V_F + V_D \, , 
\quad
V_G = - 3 \, e^G  \le 0 \, ,
$$
\begin{equation}
 \label{vgen}
V_F = e^G  G^i G_i \ge 0 \, , 
\quad
V_D =  \frac12 D_a D^a \ge 0 \, .
\end{equation}
In the above equations, $e^G$ is the field-dependent gravitino mass term $m_{3/2}^2$, $G_i = \partial G / \partial z^i$, scalar field indices are raised with the inverse K\"ahler metric $G^{i \overline{k}}$, gauge indices are raised with $[(Re f)^{-1}]^{ab}$, and
\begin{equation}
 \label{eq:solD} 
D_a = i \, G_i \, X_a^i = i \, K_i \, X_a^i \, .  
\end{equation}
For a linearly realized gauge symmetry, $i \, K_i \, X_a^i = - K_i \, (T_a)^i_{\; k} z^k$, where $T_a$ are Hermitean generators, whilst for an axionic U(1) symmetry $X_a^i = i \, q_a^i$, where $q_a^i$ is a real constant, and we obtain the so-called field-dependent Fayet--Iliopoulos terms.
Notice that D-terms are actually proportional to F-terms, $F_i = e^{G/2} \, G_i$, which implies that there cannot be pure D-breaking of supergravity in Minkowski space.

\subsection{$N=1$ no-scale models with pure F-breaking}
\label{subsec:puref}

The simplest no-scale model, originally proposed in \cite{Cremmer:1983bf}, contains just a chiral multiplet  ${\cal T} \sim (T , \widetilde{T})$, with K\"ahler potential
\begin{equation}
 \label{kaold}
K = - 3 \,  \log \left( T + \overline{T} \right) \, ,
\end{equation}
describing the non-compact $SU(1,1)/U(1)$ manifold, and a $T$-independent superpotential
\begin{equation}
 \label{wmod}
W = W_0 \, .
\end{equation}
Since $G^TG_T = 3$, $V_0=V_G + V_F =0$ and supersymmetry is broken with vanishing vacuum energy for any constant value of the massless complex scalar $T=t + i \, \tau$ ($t>0$). 
The Goldstino $\widetilde{T}$ is absorbed by the gravitino, with $m_{3/2}^2 = |W_0|^2/(8 \, t^3)$, so that $t$ plays the role of a `dilaton', setting the scale of the only non-vanishing mass term.

The model can be easily generalized to include additional chiral multiplets $\phi^{\widehat{k}}$ and vector multiplets $U^a$, as long as the equations $\langle G_{\widehat{k}} \rangle = \langle D_a \rangle = 0$ admit solutions.
For example, we can introduce a universal gauge kinetic function $f_{ab} = \delta_{ab} \, T$ and, considering small field fluctuations around $\langle z^{\hat{k}} \rangle = 0$, correct $K \to K + \Delta K$ and $W \to W + \Delta W$, with
\begin{equation}
\label{DeltaKw}
\Delta K =\sum_{\widehat{k}}  \left| z^{\widehat{k}} \right|^2 \, (T + \overline{T})^{\lambda_{\widehat{k}}} \, ,
\qquad
\Delta  W = \sum_{\widehat{k} \widehat{l}  \widehat{m}} d_{\widehat{k} \widehat{l} \widehat{m}} \,  z^{\widehat{k}} z^{\widehat{l}} z^{\widehat{m}} \, .
\end{equation}
Then the spectrum depends on other supersymmetry-breaking masses, such as gaugino masses $(m_{1/2}^2)_a= m_{3/2}^2$ and scalar masses $(m_0^2)_{\widehat{k}} = (1 + \lambda_{\widehat{k}} ) \, m_{3/2}^2$, and extra cubic scalar interactions are generated, with coefficients $(A)_{\widehat{k} \widehat{l} \widehat{m}} =  (3 + \lambda_{\widehat{k}} + \lambda_{\widehat{l}}+ \lambda_{\widehat{m}}) \, m_{3/2}$.
Neglecting for simplicity additional model-dependent discussions related with the Higgs sector, this can be taken as a starting point for obtaining realistic supersymmetric extensions of the SM, with spontaneously broken local supersymmetry.    

No-scale models can be also considered, where $n > 1$ fields $z^\alpha$ have $\langle G_\alpha \rangle \ne 0$ and take part in the exact cancellation between $V_G$ and $V_F$.
In such a case we can split the $N_T$ chiral multiplets as $\phi^i = (\phi^\alpha , \, \phi^{\widehat{k}})$, and again the no-scale properties are preserved as long as the identity $G^\alpha G_\alpha = 3$ holds and the equations $\langle G_{\widehat{k}} \rangle = \langle D_a \rangle = 0$ admit solutions.
A class of no-scale models studied in \cite{Ferrara:1994kg}, and inspired by superstring compactifications, is based on a superpotential $W(z^{\widehat{k}})$ that does not depend on the $z^{\alpha}$, and on a K\"ahler function that, expanding for small field fluctuations around $\langle z^{\widehat{k}} \rangle = 0$, reads:
\begin{equation}
\label{totkf}
K = - \log Y (r^{\alpha}) + \sum_A K^A_{\widehat{k}_A \overline{{\widehat{l}}}_A}
(r^{\alpha}) z^{\widehat{k}_A} \overline{z}^{\overline{\widehat{l}}_A}+ {1 \over 2} \sum_{A,B}
\left[ P_{\widehat{k}_A \widehat{l}_B}  (r^{\alpha}) z^{\widehat{k}_A} z^{\widehat{l}_B} + {\rm h.c.}
\right] \, .
\end{equation}
In the above equations, $Y$ is a homogeneous function of degree $3$, depending only on the combinations $r^\alpha \equiv z^\alpha + \overline{z}^{\overline{\alpha}}$.
In other words, $r^\alpha Y_\alpha = 3 \, Y$, where it is unambiguous to define $Y_\alpha \equiv \partial Y / (\partial r^{\alpha}) \equiv \partial Y / (\partial z^{\alpha}) \equiv \partial Y / (\partial \overline{z}^{\overline{\alpha}})$, and from this $G_\alpha G^\alpha =3$ follows immediately. 
The  $n_A \times n_A$ matrices $K^A_{\widehat{k}_A \overline{\widehat{l}}_A}$ are homogeneous functions of degree $\lambda_A$, i.e. $r^{\alpha} K^A_{\widehat{k}_A \overline{\widehat{l}}_A \, \alpha} = \lambda_A K^A_{\widehat{k}_A \overline{\widehat{l}}_A}$ ($\sum_A n_A = N_T - n$).
The functions~$P_{\widehat{k}_A \widehat{l}_B}$ have analogous scaling properties, $r^{\alpha} P_{\widehat{k}_A \widehat{l}_B \, \alpha} = \rho_{\widehat{k}_A \widehat{l}_B} P_{\widehat{k}_A \widehat{l}_B}$.
We also assume that the gauge field metric, ${\rm Re \,}f_{ab}$, is a homogeneous function of the variables $r^{\alpha}$ of degree $\lambda_f$, i.e. $r^{\alpha} ({\rm Re \,} f_{ab})_{\alpha} = \lambda_f  \, {\rm Re \,} f_{ab}$  ($\lambda_f=0,1$).
For the models under consideration, the general supergravity mass formulae undergo dramatic simplifications.
The spin 0 fields $z^{\alpha}$ in the supersymmetry breaking sector have classically vanishing masses, therefore their masses will be induced by quantum corrections.
After subtracting the goldstino, their spin $1/2$ partners $\psi^{\alpha}$ have all masses equal to the gravitino mass $m_{3/2}$.
Furthermore, remembering that the chiral superfields $z^{\widehat{k}}$ should contain the quark, lepton and Higgs superfields of the Minimal Supersymmetric extension of the Standard Model (MSSM), we can derive some predictions for the explicit mass parameters of the MSSM. 
Similar predictions were derived in \cite{Brignole:1993dj}, for special goldstino directions and under slightly different assumptions.
For the gaugino masses we find that, if there is unification of the gauge couplings, $({\rm Re \,} f)_{ab} = \delta_{ab} / g_U^2$, then $m_{1/2}^2 = \lambda_f^2 \, m_{3/2}^2$ $(\lambda_f=0,1)$.
As for the spin $1/2$ fermions $\psi^{\widehat{k}}$, we should distinguish two main possibilities. 
Those in chiral representations of the gauge group, such as quarks and leptons, cannot have gauge-invariant mass terms.
Those in real representations of the gauge group, such as the Higgsino fields $\widetilde{H}_1$ and $\widetilde{H}_2$ of the MSSM, can have both a `superpotential mass', proportional to $W_{\widehat{k}_A \widehat{l}_B}$, and a `gravitational' mass, proportional to $P_{\widehat{k}_A \widehat{l}_B}$, in their mass terms $(M_{1/2})_{\widehat{k}_A \widehat{l}_B}$. 
Both these terms can in principle contribute, in a suitable flat limit, to the superpotential `$\mu$-term' of the MSSM, and to the associated off-diagonal (analytic-analytic) scalar mass term $m_3^2$, but in many examples either the superpotential or the gravitational contribution to $\mu$ will be present, not both.
Writing then $(M_0^2)_{\widehat{k}_A \widehat{l}_B} = (B)_{\widehat{k}_A \widehat{l}_B} (M_{1/2})_{\widehat{k}_A \widehat{l}_B}$, in analogy with the MSSM notation, we find
$B_{H_1 H_2} = (2 + \lambda_{H_1} + \lambda_{H_2}) m_{3/2}$
or
$
B_{H_1 H_2} = (2 + \lambda_{H_1} + \lambda_{H_2} - \rho_{H_1 H_2}) m_{3/2}$,
respectively. 
Moving further to the spin 0 bosons $z^{\widehat{k}}$ in chiral representations (squarks, sleptons, \ldots), they can only have diagonal (analytic-antianalytic) mass terms, of the form $(m_0^2)_A = (1+\lambda_A) \, m_{3/2}^2$.
In contrast with quarks and leptons, scaling weights smaller than $(-1)$ are allowed for the Higgs fields, since in that case a negative contribution to $m_0^2$ can be compensated by an extra positive contribution coming from the $\mu$-term. 
Similarly, a general formula can be obtained for the coefficients of the cubic scalar couplings of the MSSM potential,
$
(A)_{\widehat{k}_A \widehat{l}_B \widehat{m}_D} = (3 + \lambda_A + \lambda_B + \lambda_D) \, m_{3/2}
$.

\subsection{$N=1$ variations with F- and D-breaking}

Are pure F-breaking and at least a complex flat direction in field space unavoidable features of $N=1$ no-scale models? 
This question was answered in the negative in \cite{DallAgata:2013jtw}, where a no-scale model was introduced with F- and D-breaking of supersymmetry and a single real massless scalar in the classical spectrum. 
The model contains a vector multiplet $U  \sim (\lambda, V_\mu)$ and a  chiral multiplet ${\cal T} \sim (T , \widetilde{T})$, with K\"ahler potential
\begin{equation}
 \label{kamod}
K = - 2 \,  \log \left( T + \overline{T} \right) \, .
\end{equation}
In contrast with the model of \cite{Cremmer:1983bf}, $\tau$ is now an `axion' that shifts under the $\widetilde{U(1)}$ isometry gauged by the vector multiplet. 
The corresponding holomorphic Killing vector is just an imaginary constant, and it is not restrictive to set the charge $q=1$ and write:
\begin{equation}
 \label{kimod}
X^T = i  \, .
\end{equation}
The most general superpotential invariant under the gauged $\widetilde{U(1)}$ is then a constant with respect to $T$, as in (\ref{wmod}).
The gauge kinetic function is a positive real constant
\begin{equation}
 \label{fmod}
f = \frac{1}{\widetilde{g}^2} \, .
\end{equation}
The scalar potential of (\ref{vgen}) is then the sum of
\begin{equation}
 \label{vgfdmod}
V_G = -  \frac{3 \, |W_0|^2}{4 \, t^2} \, , 
\qquad
V_F =  \frac{|W_0|^2}{2 \, t^2} \, , 
\qquad
V_D = \frac{\widetilde{g}^2}{2 \, t^2} \, .
\end{equation}
As required by gauge invariance, $V_0$ does not depend on $\tau$: the axion is absorbed by the massive $\widetilde{U(1)}$ vector boson via the Higgs effect. 
However, $V_G$, $V_F$ and $V_D$ all depend non-trivially on $t$.
Constant $W$, constant $f$ and the $(-2)$ factor multiplying the logarithm in $K$ are essential in ensuring that all three terms in (\ref{vgfdmod}) have the same $t$-dependence.
In particular, for
\begin{equation}
 \label{choice}
|W_0| = \sqrt{2} \, \widetilde{g} \, , 
\end{equation}
there is an exact cancellation and $V_0=V_G+V_F+V_D = 0$. 
The gauge symmetry and supersymmetry are broken on Minkowski space at all classical vacua, with the would-be Goldstone boson and fermion given by $\tau$ and by a linear combination of $\widetilde{T}$ and $\lambda$, respectively. 
The only classically massless particle is the real scalar $t$, all the other particles in the physical spectrum have squared masses proportional to $m_{3/2}^2  = \widetilde{g}^2  / (2 \, t^2)$.
Superficially, we may think that the choice of (\ref{choice}) is a fine-tuning, but it can be shown \cite{DallAgata:2013jtw}  that such term can be originated by a consistent $N=1$ truncation of a $N=2$ model with a single gauge interaction.
Similarly to the model in \cite{Cremmer:1983bf}, also this model can be straightforwardly generalised to include additional vector and chiral multiplets, with gauge group $\widetilde{U(1)}\times {\cal G}$, as long as the latter do not transform under the original $\widetilde{U(1)}$.

If $m_{3/2}$ and $m_{W}$ are both to be dynamically determined by dimensional transmutation, irrespectively of their ratio, we may ask whether there are no-scale models where supersymmetry and the electroweak gauge symmetry are both spontaneously broken, with a positive semidefinite classical potential and both scales sliding along classically flat directions in field space. 
The answer is positive, and some examples have been given in \cite{Brignole:1994zz,Brignole:1995cq,Luo:2014uha}.
We briefly describe here, for illustration, the model introduced in \cite{Luo:2014uha}, which extends the one in \cite{DallAgata:2013jtw} and has  only two real classical flat directions, in one-to-one correspondence with the scales of supersymmetry and electroweak symmetry breaking. 
In short, the  model couples the hidden sector of \cite{DallAgata:2013jtw} to the electroweak gauge and Higgs sector of the MSSM. 
The gauge group is $SU(2)_L \times U(1)_Y \times \widetilde{U(1)}$.
The chiral multiplets are the SM-singlet ${\cal T} \sim (T,\widetilde{T})$ and the two MSSM Higgs doublets, ${\cal H}_1 \sim (H_1, \widetilde{H}_1)$ and  ${\cal H}_2 \sim (H_2, \widetilde{H}_2)$. 
Under $\widetilde{U(1)}$, the imaginary part of $T$ shifts as before, whilst the two Higgs superfields do not transform.
Motivated by string compactifications and by extended supergravities, we choose the K\"ahler manifold for the scalar fields (unifying the chiral multiplets in the hidden and Higgs sectors) to be $SO(2,5)/[SO(2) \times SO(5)]$:
\begin{equation}
e^{-K} = ( T + \overline{T})^2 - | H_1^0 - \overline {H_2^0} |^2 - |H_1^- + \overline{H_2^+} |^2 \, ,
\label{kahler}
\end{equation}
and, in the field basis of (\ref{kahler}), the constant superpotential of (\ref{choice}).
Finally, we choose a factorised gauge kinetic function:
\begin{equation}
\widetilde{f} = \frac{1}{ \widetilde{g}^{\, 2}} \, , 
\qquad
f_Y = a_Y + b_Y \, T \, , 
\qquad
f_L = a_L + b_L \, T \, ,
\label{gkinf}
\end{equation}
where $(\widetilde{g},a_Y,b_Y,a_L,b_L)$ are real constants. 
The classical potential $V_0$ turns out to be the sum of positive semidefinite contributions proportional to $\widetilde{g}^2$ and to the field-dependent $SU(2)_L$ and $U(1)_Y$ coupling constants $ g^{\, \prime \, 2} \equiv 1/(Re \, f_Y)$ and  $g^2 \equiv 1/(Re \, f_L)$.
After gauge fixing, inequivalent vacua can be classified by $\langle H_1^- \rangle = \langle H_2^+ \rangle = 0$ and
\begin{equation}
\label{vevs}
\langle T \rangle = x \, , 
\qquad \qquad
\langle H_1^0 \rangle = \langle H_2^0 \rangle = 2 \, x \, v \, ,
\end{equation}
where $x>0$ and $v \ge 0$ parametrise two real flat directions. As in \cite{DallAgata:2013jtw}, the $\widetilde{U(1)}$ gauge symmetry and supersymmetry are spontaneously broken on flat Minkowski space at all vacua. The electroweak gauge symmetry is also spontaneously broken on the generic vacuum, although it can be restored at the special point $v=0$. 
In the hidden sector, the spectrum is exactly as in \cite{DallAgata:2013jtw}.
Setting here $T = x \, (1 + t + i \, \tau)$, $\tau$ is the Goldstone boson absorbed by the massive $\widetilde{U(1)}$ vector, and $t$ is a canonically normalised massless scalar. 
In the observable sector, the spectrum of gauge and Higgs bosons, charginos and neutralinos corresponds to a special choice of parameters in the MSSM. 
The only classically massless fields are the dilaton $t$ in the hidden sector and the SM Higgs boson $h$ in the MSSM Higgs sector~\footnote{A classically massless $h$ is not a (qualitative) problem: if quantum corrections eventually select $v$, a mass for $h$ is automatically generated.}, where the masses of all exotic scalars receive supersymmetry-breaking contributions of order $m_{3/2}$. 
All renormalisable interactions are exactly as in the MSSM, for a special parameter choice, and the non-renormalisable interactions are suppressed by inverse powers of $M_P$.
Finally, the matter sector and the strong interactions of the MSSM can be included in the observable sector in a straightforward way, although at the cost of some additional model dependence. 

\subsection{$N=1$ non-linear realisations and inflation}

No-scale models can be combined with non-linear realisations of supersymmetry to build semi-realistic inflationary models where, at the end of inflation, supersymmetry is spontaneously broken with naturally vanishing classical vacuum energy and no classically massless scalars in the spectrum.
A review of inflationary models making use of non-linear supergravity and an extensive list of references can be found in \cite{Antoniadis:2024hvw}.
Here  \cite{DallAgata:2014qsj} we recall first how, starting from eqs.~(\ref{kaold}) and (\ref{wmod}), we can move to a model where the Goldstino belongs to a nilpotent chiral superfield $S$, its complex scalar partner $s$ ({\em sgoldstino}) is removed from the spectrum, the gravitino mass is fixed and the classical vacuum energy naturally vanishes.
Then we comment on how this model can be extended to include the additional unconstrained chiral multiplets of a realistic inflationary model, in particular the inflaton $\Phi$.

After the analytical field redefinition $Z=(2T-1)/(2T+1)$, the model of eqs.~(\ref{kaold}) and (\ref{wmod}) is equivalent to the one defined by
\begin{equation}
 \label{kwredef}
K = - 3 \,  \log \left( 1 - |Z|^2 \right) \, ,
\qquad \qquad
W=W_0 \, (1-Z)^3 \, .
\end{equation}
Making the additional assumption that the superfield $Z$ is nilpotent, $Z^2=0$, expanding around $Z = 0$ we get
\begin{equation}
K = 3 |Z|^2 \, , 
\qquad \qquad
W = W_0 \, (1 - 3 Z) \, .
\end{equation}
A simple constant rescaling,  $Z = - S/\sqrt{3}$, gives then
\begin{equation}
\label{finalW}
K = |S|^2 \, , 
\qquad \qquad
W = W_0 \, (1 + \sqrt3 \, S) \, .
\end{equation}
It is then immediate to check that, for $\langle s \rangle =0$ as required by the nilpotency condition, $\langle F^S \rangle = \sqrt3\, \overline{W_0} \ne 0$. 
This implies that, as in the no-scale model considered above, supersymmetry is spontaneously broken with classically vanishing vacuum energy.
However, now the gravitino mass is fixed to $m_{3/2}^2 = |W_0|^2$ and there are no classically flat directions: $S^2=0$ implies $\langle s \rangle =0$ and the sgoldstino $s$ is removed from the spectrum.    

Adding an unconstrained chiral multiplet $\Phi$, whose complex scalar $\phi = (a + i \, \varphi)/\sqrt{2}$ contains the inflaton $\varphi$ and its scalar partner $a$,  it is easy to build models that allow for slow-roll, large field inflation with $a$ frozen at $\langle a \rangle = 0$ by a large $\varphi$-dependent mass during the inflationary phase \cite{DallAgata:2014qsj}:
\begin{equation}
K = \frac12 (\Phi + \bar \Phi)^2 + |S|^2 \, , 
\qquad
\qquad
W = f(\Phi) \, (1 + \sqrt3\, S) \, ,
\end{equation}
where $f(z)$ is a holomorphic function with the following properties:
\begin{equation}
\label{fcond}
\overline{f(z)} = f(- \overline{z}) \, , 
\qquad
\qquad
f^{\, \prime} ( 0 ) = 0 \, , 
\qquad
\qquad
f ( 0 ) \ne 0 \, .
\end{equation}
The above models can be also coupled to the `visible' matter sector of supersymmetric extensions of the SM without spoiling the nice features of the inflationary potential. 

\subsection{$N>1$ no-scale models}

With respect to simple $N=1$, $d=4$ supergravity, extended $N>1$ supergravities have the fatal phenomenological flaw of not admitting chiral fermions, but the theoretical advantage of being more and more constrained as $N$ increases up to its maximal value $N=8$.
Therefore, they are important theoretical laboratories to understand, for example, supersymmetry breaking at the classical and at the quantum level.  
It is a remarkable empirical fact, certainly deserving a deeper understanding, that all known $N > 1$ supergravity models with spontaneously broken supersymmetry and classically vanishing vacuum energy are no-scale models.  
In the following, for simplicity, we will consider only the cases $N=2,4,8$.

Whilst the gravitational multiplet exhausts the field content of $N=8$, $N=4$ can contain vector multiplets, and $N=2$ can contain vector multiplets and hypermultiplets. 
The manifold of the scalar fields is $E_7/SU(8)$ for $N=8$, $SU(1,1)/U(1) \times SO(6,n)/[SO(6) \times SO(n)]$ for $N=4$, and the product of a special K\"ahler manifold and a quaternionic K\"ahler manifold for $N=2$.
The only way to introduce a non-trivial scalar potential and the spontaneous breaking of supersymmetry is the {\em gauging} procedure (for recent reviews and references, see e.g.
\cite{Trigiante:2016mnt,DallAgata:2021uvl,Sezgin:2023hkc}), which consists in promoting a subgroup of the global isometry group of the scalar manifold to a local symmetry. 
The non-linear action of the global isometries of the scalar manifold is associated with an electric/magnetic duality action on the vector field strengths and their duals \cite{Gaillard:1981rj}, and thus is defined by the embedding inside $Sp(2 n)$, if $n$ is the number of vector fields. 
The general form of the scalar potential $V_0$, which extends (\ref{vgen}) to $N>1$, follows from \cite{ferraramaiani,DAuria:2001rlt}:
 \begin{equation}
 \delta_B^A \,V_0 = - 3 \, S^{AC} \,S_{BC} + N^{IA} \,N_{IB} \, ,
\end{equation}
where 
$\delta \psi_{A \mu} = S_{AB} \gamma _\mu \varepsilon ^B + \ldots$ 
and
$\delta \lambda^I = N^{IA} \varepsilon_A + \ldots$
are the variations of the gravitinos and of the spin-1/2 fermions under supersymmetry transformations, with $N_{IA} = (N^{IA})^*$ and $S_{AB} = S_{BA} = (S^{AB})^*$.
As in the $N=1$ case, $V_0$ consists of two contributions: one is positive semidefinite and corresponds to the sum of the squares of the would-be auxiliary fields for the non-gravitational multiplets, the other is negative semidefinite and proportional to the  squared gravitino mass matrix.

No-scale models exhibiting complete spontaneous breaking of extended supesymmetry with vanishing classical vacuum energy can be associated, in $N=2$ \cite{Gunaydin:1984pf,Cremmer:1984hj}, $N=4$ \cite{deRoo:1985jh,Tsokur:1994gr,Villadoro:2004ci}  and $N=8$ \cite{Cremmer:1979uq, Andrianopoli:2002mf, DallAgata:2011aa} supergravity, with the gauging of a non-compact subgroup of the duality group.
If, in addition, for a subset of fields $\lambda^{\widetilde{I}}$
\begin{equation}
\sum_C 3 \, S^{AC}\,S_{CA} = \sum_{\widetilde{I}}  N^{\widetilde{I}A}\,N_{\widetilde{I}A}
\qquad
\forall A
\end{equation}
along some flat scalar directions that control the scale of the gravitino masses, then the residual potential
$
\sum_{I \ne \widetilde{I}}  N_I^A \,N^I_A
$
is manifestly positive semidefinite and we fall into a more stringent definition of no-scale model.

\subsection{Quantum stability and hierarchies}

In no-scale models, the obvious first tool for exploring the approximate stability of the classical Minkowski background and the dynamical generation of hierarchies is the one-loop effective potential $V_1=V_0+\Delta V_1$, computed as a function of those background fields whose constant value is left undetermined at the classical level.   
$\Delta V_1$ is in general divergent, and using a momentum cut-off $\Lambda$ reads  \cite{Coleman:1973jx, Weinberg:1973ua} 
\begin{equation}
\label{veff}
\Delta V_1 = 
{{\rm Str \,} {\cal M}^0 \over 64 \pi^2}  \, \Lambda^4 \log \Lambda^2
+ 
{{\rm Str \,} {\cal M}^2 \over 32 \pi^2} \, \Lambda^2 
+ 
{1 \over 64 \pi^2} {\rm Str \,} {\cal M}^4 \log {{\cal M}^2 \over \Lambda^2}  
\, .
\end{equation}
Its ultraviolet behaviour is controlled by the supertraces of the even powers of the field-dependent mass matrices, defined by 
\begin{equation}
{\rm Str} \, {\cal M}^{2k}  \equiv  \sum_I  (-1)^{2J_I} \,  (2J_I +1) \, (M^2_I)^k 
\label{Supertrace}
\end{equation}
where $k=0,1,2,\ldots$, the index $I$ runs over the different particles in the spectrum, $M^2_I$  and $J_I$ are the corresponding squared-mass eigenvalues and spins.  Of course, massless particles do not contribute to the supertraces. 

The (field-independent) quartic divergence, which would be present in a generic $d=4$ theory, is proportional to $Str \, {\cal M}^0$, and is always absent in theories with spontaneously broken supersymmetry, where the number of bosonic and fermionic degrees of freedom is the same, $n_B=n_F$.  
Theories with spontaneously broken $N=1$, $d=4$ supersymmetry have in general a quadratically divergent contribution to $V_{1}$ proportional to ${\rm Str} \, {\cal M}^2$.
Only in theories with spontaneously broken $N=4$, $d=4$ supergravity it is guaranteed \cite{Grisaru:1982sr,DallAgata:2023ahj} that ${\rm Str} \, {\cal M}^2=0$, so that the divergent contribution to $V_{1}$ is only logarithmic and proportional to ${\rm Str} \, {\cal M}^4$. 
Theories with spontaneously broken $N=8$, $d=4$ supergravity have the special property that ${\rm Str} \, {\cal M}^2= {\rm Str} \, {\cal M}^4 = 0$ and $V_{1}$ is finite \cite{Cremmer:1979uq,DallAgata:2012tne}.
This makes the one-loop effective potential calculable, for all tachyon-free constant field configurations along the classically flat directions,
\begin{equation}
\label{respot}
V_1 = \frac{1}{64 \, \pi^2} \, {\rm Str} \, \left( {\cal M}^4 \, \log {\cal M}^2 \right) \, .
\end{equation}
In all cases studied so far, it was found \cite{Cremmer:1979uq,DallAgata:2012tne,Catino:2013ppa} that $ {\rm Str} \, {\cal M}^6=0$, $ {\rm Str} \, {\cal M}^8 >0$ and $V_1<0$, so that at the one-loop level no locally stable vacua are known with fully broken supersymmetry and positive or vanishing vacuum energy.
This last result is somewhat disappointing, but as we will discuss later it cannot be considered to be the end of the story. 

In the case of $N=1$ no-scale models, it is useful to elaborate more on ${\rm Str} {\cal M}^2$, the coefficient of the one-loop quadratically divergent contributions to the vacuum energy.
Assuming for simplicity pure F-breaking,  we can write in general \cite{Grisaru:1982sr}
\begin{equation}
\label{genstr}
{\rm Str \,} {\cal M}^2 (z, \overline{z}) = 2 \, Q (z, \overline{z}) \, m_{3/2}^2 (z, \overline{z})
\, ,
\end{equation}
where
\begin{equation}
\label{qexpr}
Q (z, \overline{z})  =  N_{T} - 1 - G^i  (z, \overline{z}) \left[ R_{i \overline{k}} (z, \overline{z}) + F_{i \overline{k}} (z, \overline{z}) \right] G^{\overline{k}} (z, \overline{z})
\end{equation}
\begin{equation}
\label{ricci}
R_{i \overline{k}} (z, \overline{z}) \equiv \partial_{i} \partial_{\overline{k}}
\log \det G_{m \overline{n}} (z, \overline{z}) \, ,
\end{equation}
\begin{equation}
\label{ficci}
F_{i \overline{k}} (z, \overline{z}) \equiv - \partial_{i} \partial_{\overline{k}}
\log \det {\rm Re \,} [ f_{ab} (z) ] \, .
\end{equation}
In eq.~(\ref{ricci}), $R_{i \overline{k}}$ is the Ricci tensor constructed from the metric of the K\"ahler manifold for the $N_T$ chiral multiplets, and $F_{i \overline{k}}$ has a similar geometrical interpretation in terms of the metric for the gauge superfields.
It is important to observe that both $R_{i \overline{k}}$ and $F_{i \overline{k}} $ {\em do not depend at all} on the superpotential of the theory. 
This very fact allows for the possibility that, for special geometrical properties of these two metrics, the dimensionless quantity $Q (z, \overline{z})$ may turn out to be field-independent.
In the models considered at the end of section \ref{subsec:puref}, for example   \cite{Ferrara:1994kg} 
\begin{equation}
\label{qfinal}
Q = \sum_A \left( 1+\lambda_A \right) n_A - n - \lambda_f d_f - 1 \, ,
\end{equation}
where $d_f$ is the dimension of the gauge group.
Requiring $Q=0$ is too strong a constraint, since the ultraviolet completion of the model, where the cutoff $\Lambda$ is to be replaced by some heavy mass scale, may induce contributions to $V_1$ of the same order, coming from heavier states whose supersymmetry-breaking mass splittings are also controlled by $m_{3/2}$.
However, the field-independence of $Q$ leaves the door open for a possible cancellation in the full theory.
%
%
%
%
%
\section{No-scale models from higher dimensions}
\label{sec:hd}

\subsection{Dimensional reductions and consistent truncations}
\label{sec:dimred}

A deeper understanding of $d=4$ no-scale supergravity can be gained by looking at those models that can be linked with supergravity or superstring theories formulated in $d>4$ dimensions.
For a long time, the only available examples were the generalised dimensional reductions of $d>4$ supergravities, introduced in \cite{Cremmer:1979uq,Scherk:1979zr} and reviewed by John Schwarz in this volume \cite{Schwarz}, which can be implemented in two different versions.
In cases such as the single extra dimension compactified on a circle considered in \cite{Cremmer:1979uq}, the twist of the boundary conditions leading to supersymmetry breaking corresponds to a global continuous R-symmetry of the higher-dimensional action: the tree-level spectrum depends on a single modulus $R$, the compactification radius, and we can smoothly take the limit of small twist parameters, corresponding to $m_{3/2} \ll M_C \sim 1/R$. 
However, when the compactification involves three or more internal dimensions we can use general coordinate transformations to twist boundary conditions \cite{Scherk:1979zr}.
This kind of Scherk--Schwarz compactifications are also known as twisted tori compactifications, since the boundary conditions act on the geometry of the internal manifold, modifying its structure.
After a field redefinition, twisted boundary conditions can be replaced by a VEV for the spin connection, which leads to observable effects because the compact space is non-simply connected, in analogy with gauge symmetry breaking by Wilson lines.
As explained for example in \cite{Kaloper:1999yr}, these VEVs can be seen as `metric fluxes', with quantized twist parameters and $m_{3/2} \sim M_C \sim 1/R$.  
The reduced theory is then a mere consistent truncation rather than a genuine effective theory \cite{DallAgata:2005zlf, Grana:2013ila, GDAFZ2025}: this is obvious when the compactification manifold remains a torus and supersymmetry appears to be broken in the reduced truncated theory but is left unbroken in the full compactified theory; however, this can also be the case when supersymmetry is broken, since in the full compactified theory there can be physical states truncated away and others included in the reduced theory with comparable masses.
 
After the introduction of compactifications of the heterotic superstring, or of the associated $N=4$, $d=10$ supergravity, preserving $N=1$, $d=4$ supersymmetry, for example on Calabi-Yau manifolds \cite{Candelas:1985en} or on six-dimensional toroidal orbifolds \cite{Dixon:1985jw}, it was soon realised that the reduced theories, containing only the finite number of massless degrees of freedom for unbroken supersymmetry, have many of the features of no-scale models \cite{Witten:1985xb,Ferrara:1986qn,Antoniadis:1987zk,Ferrara:1987tp,Ferrara:1987jr}.
The only missing ingredient is a suitable superpotential triggering supersymmetry breaking, which of course cannot be derived from compactifications that preserve $N=1$ supersymmetry by construction.
In the first superstring-inspired versions of $N=1$, $d=4$ no-scale models, this superpotential was introduced as an ad hoc assumption, motivated by non-perturbative effects \cite{Dine:1985rz, Ferrara:1987jq}.
The first no-scale models actually derived from coordinate-dependent compactifications of the heterotic superstring \cite{Rohm:1983aq, Ferrara:1987es, Kounnas:1988ye}  \`a la Scherk-Schwarz were obtained in \cite{Ferrara:1988jx, Porrati:1989jk}.

With the advent of flux compactifications of superstring theories, many new possibilities for generating $d=4$ no-scale models opened up \cite{Giddings:2001yu, Kachru:2002he,  Derendinger:2004jn,  Villadoro:2005cu, Camara:2005dc}.
In the following, for illustration, we briefly summarize the $N=1$ options that arise when compactifying the heterotic, Type IIA and Type IIB superstring theories on the orbifold $T^6/(Z_2 \times Z_2)$, with an additional $Z_2$ orientifold projection in the case of the $N=8$ Type II theories. 
For simplicity, we ignore matter fields and concentrate on the chiral superfields containing the seven main moduli $(z^k)_{k=1,\ldots,7} = (S,T_1,T_2,T_3,U_1,U_2,U_3)$, where $S$ is usually called dilaton, $T_A$ K\"ahler moduli and $U_A$ complex structure moduli ($A=1,2,3$). 
Their real parts come from the dilaton and the diagonal components of the internal metric, their imaginary parts come from internal components of the $p$-forms and from the off-diagonal components of the internal metric.
Although their identification in terms of the $d=10$ bulk fields is model-dependent, they are described by a common K\"ahler potential with well-defined scaling properties
\begin{equation}
K = - \sum_{k=1}^7 \log (z^k + \overline{z}^{\overline{k}}) \, ,
\qquad \qquad
\sum_{k=1}^7 K^k \, K_k = 7  \, .
\end{equation}
We then expect, on the basis of the general discussion in section~\ref{subsec:puref}, that no-scale models can be generated if $W$ has a non-trivial dependence on four moduli $(z^{\widehat{k}})$ and does not depend on three of them $(z^\alpha)$.
The origin of $W$ resides, at the perturbative level, in the possible geometrical fluxes that can be turned on in the different models: these fluxes include the $(p+1)$-form field strengths of the $p$-forms present in the model, as well as metric fluxes (we neglect here non-geometrical fluxes, which can be introduced by exploiting the correspondence between fluxes and the gauging of the duality symmetries of extended supergravities).
Indeed, it can be shown that the most general $N=1$ superpotential generated by fluxes is a polynomial of the form 
\begin{equation}
\label{wfluxes}
W = a + i \, a_k  \, z^k + a_{kl} \, z^k z^l + i \, a_{klm} \, z^k z^l z^m + \ldots \, , 
\end{equation}
where the real coefficients $(a,a_k,a_{kl},a_{klm},\ldots)$ ($k<l<m<\ldots=1,\ldots,7$) are constrained by the possible fluxes that are compatible with the bosonic field content of the model, with the orbifold and orientifold projections and with the Bianchi identities (BI) of local symmetries gauged by either bulk or brane-localized vectors.
In the heterotic case, the available fluxes are 24 metric fluxes $\omega_3$ and 8 fluxes $H_3$ of the NS-NS 3-form, subject to quadratic constraints of the generic form $\omega_3 \cdot \omega_3 =0$ and $\omega_3 \cdot H_3 = 0$: the former induce in $W$ terms containing  1 $T_A$ and $0/1/2/3$ $U_A$, the latter induce terms with no $S$, no $T_A$ and $0/1/2/3$ $U_A$.   
In this case, there are only bulk vector fields and the BI are equivalent to the Jacobi identities of $N=4$ gauged supergravity \cite{Kaloper:1999yr}.  
A simple example of no-scale model generated by metric fluxes only is obtained with $W= k (T_2 U_2 + T_3 U_3)$. 
In the Type-IIA case, where O6-planes and D6-branes (O6/D6) are also present, the bulk fluxes compatible with the orbifold and orientifold projections are: 
12 $\omega_3$, 4 $H_3$ and (1+3+3+1) fluxes for the RR field strengths $(F_0,F_2,F_4,F_6)$. 
In this case, the discussion of the BI needs to take into account the contributions from localized sources, and is no longer equivalent to the Jacobi identities of a $N=4$ gauging \cite{Villadoro:2005cu}. 
The rich system of fluxes can induce in $W$ dependences on all 7 main moduli, but at most linear in the $U_A$. 
Several options for no-scale models arise, for example:  
$W=a(T_1U_1+T_2U_2)$, purely from $\omega_3$;
$W=a(ST_1+T_2T_3) + i b (S+T_1T_2T_3)$, from $(\omega_3,F_2,H_3,F_6)$;
$W=a(ST_1+ST_2+ST_3+T_1T_2+T_2T_3+T_3T_1) +3 i b (S+T_1T_2T_3)$, from $(\omega_3,F_0,H_3,F_2)$.
In the Type-IIB case, one can choose either (O3/D3+O7/D7) or  (O5/D5+O9/D9) systems and proceed accordingly.
For example, with (O3/D3+O7/D7)  we can consider $H_3$ and $F_3$ fluxes, constrained by suitable BI involving localized sources, and generate $W(S,U_1,U_2,U_3)$ with no dependence on $(T_1,T_2,T_3)$: again, there are acceptable choices leading to no-scale models. 

All of the above can be generalized to include the scalars coming from the untwisted sectors, and in some cases also the twisted sectors.

\subsection{Full-fledged compactifications}

The behaviour of a reduced $d=4$ supergravity theory does not capture the physics of the full compactified theory.
In the case of no-scale supergravity, such theory contains also infinite towers of Kaluza--Klein (KK) modes, with supersymmetry-breaking mass splittings that depend on the classically massless moduli, associated with geometrical properties of the compactification manifold. 
In addition, compactified string theories also contain winding modes, corresponding to strings wrapped around the compactification manifold, whose effects exponentially decouple in the
$M_C \ll M_S$ limit.  
String models with localized defects such as orbifold fixed points, D-branes, O-planes, etc give rise to additional non-perturbative states in the spectrum and make the effective supergravity more complicated to analyse and not always under full control. 

In the rest of this section, we consider $d>4$ no-scale supergravities where supersymmetry is broken in a Scherk-Schwarz compactification, to understand some important qualitative differences with respect to their reduced $d=4$ versions, especially for what concerns the quantum corrections along the flat directions of the classical potential. 
We expect our considerations to be valid also for their string theory completions, when they exist, as long as they do not involve localized defects and the limit $M_C \ll M_S$ can be taken.
Our goal is to show that:   
the full four-dimensional one-loop effective potential $V_1$ and its counterpart $V_{1,red}$ evaluated in the reduced theory can drastically differ;
the non-locality of supersymmetry breaking \`a la Scherk--Schwarz (or, more generally, by fluxes) guarantees \cite{Rohm:1983aq} the finiteness of $V_1$;
the KK states are organized in $N > 1$ multiplets with supersymmetry-breaking mass splittings.

As a first pedagogical example  \cite{DallAgata:2024ijh}, we consider  pure $N=2,4,8$, $d=5$ supergravities with complete supersymmetry breaking in the Scherk-Schwarz compactification to $d=4$, compute  $V_1$, express it as a function of the radial modulus and of the twist parameters, discuss the precise correspondence between the finite $V_1$ and $V_{1,red}$  (divergent for $N=2,4$ and finite for $N=8$).
For Scherk--Schwarz compactifications on a circle:
\begin{equation}
V_1 = \frac{1}{2} \, 
\int \frac{d^4p}{(2 \pi)^4} \, 
\sum_{n=- \infty}^{+ \infty} \, 
\sum_I \, 
(-1)^{2 J_I} \,
(2 J_I + 1) \,
\log \left( p^2 + m_{n,I}^2 \right)  \, ,
\label{V1gen}
\end{equation}
where the index $I$  runs over the finite number of independent $d=4$ modes of spin $J_I$, with $(2 J_I+1)$ degrees of freedom (dof) and field-dependent mass 
\begin{equation}
m_{n,I}^2 =
\frac{(n+s_I)^2}{R^2}
\qquad
\qquad
\left(
n = 0, \pm1, \pm2, \ldots
\right) 
\label{mass}
\end{equation}
at each of the infinite KK levels $n \in \mathbf{Z}$. 
Here $R$ is the physical field-dependent radius of the circle, expressed in units of $M_P$, and the shifts $s_I$ in the mass formula (\ref{mass}) are determined by the Scherk--Schwarz twists.
Since supersymmetry breaking is non-local in the compact dimension and supersymmetric $d=4$ Minkowski vacua are perturbatively stable, $V_1$ is finite \cite{Rohm:1983aq} and its explicit calculation gives
\begin{equation}
V_1 = - \frac{3 }{128 \, \pi^6 \, R^4}  \,
\sum_I \, 
(-1)^{2 J_I} \,
(2 J_I + 1) \,
\left[ {\rm Li_5} (e^{- 2 \pi \, i \, s_I}) +  {\rm Li_5} (e^{2 \pi \, i \, s_I})  \right] 
\, ,
\label{V1res}
\end{equation}
where ${\rm Li_n} (x)=\sum_{k=1}^{\infty} x^k/k^n$ are polylogarithms and ${\rm Li_5}(1)=\zeta(5)\simeq 1.037$.

It is interesting to understand in some detail the relation between $V_1$ and $V_{1,red}$.
The latter has an equal and finite number of bosonic and fermionic degrees of freedom, with masses given by Eq.~(\ref{mass}) for $n=0$.
We expect the reduced theory to make sense for $|s_I | \ll 1$, leading to $m_{0,I}^2 \ll (1/R^2)$, with an effective cutoff $\Lambda \sim 1/R$. 
We start by expressing $V_{1,red}$ as in (\ref{veff}), in terms of the mass matrix ${\cal M}_0^2$ for the $n=0$ KK level, where in general  
\begin{equation}
 {\rm Str} \, {\cal M}_n^p 
 \equiv
 \sum_I \, 
(-1)^{2 J_I} \,
(2 J_I + 1) \,
m_{n,I}^p
\, .
\label{strdef}
\end{equation}
We observe that, for $|s_I| \ll 1$:
\begin{eqnarray}
\left[ {\rm Li_5} (e^{- 2 \pi \, i \, s_I}) +  {\rm Li_5} (e^{2 \pi \, i \, s_I})  \right] 
& = &
2 \, \zeta(5) 
- 4 \pi^2 \, \zeta(3) \, s_I^2
+ \frac{\pi^4}{3} \left[ \frac{25}{3} - 4 \, \log (2 \pi) \right] \, s_I^4
\nonumber \\ & - & 
\frac{2}{3} \, \pi^4 \, s_I^4 \, \log s_I^2
+ \ldots
\, ,
\label{V1exp}
\end{eqnarray}
where the first term $2 \, \zeta(5)$ can be ignored, since it does not survive the supertrace, and  the dots stand for terms of order $s_I^6$ and higher. 
In the reduced theory, ${\rm Str} \, {\cal M}_0^2 = {\rm Str} \, s_I^2 /R^2$. 
Therefore, if $ {\rm Str} \, s_I^2 \ne 0$, the leading contributions to the quadratically divergent $V_{1,red}$ and to the finite $V_1$ are those of order $s_I^2$. 
Equating the two, the effective cutoff for the reduced theory is
$\Lambda = \sqrt{3 \, \zeta(3)}/(\pi \, R) \simeq 0.6/R$,
in agreement with the expectations.
Suppose now that   ${\rm Str} \, {\cal M}_0^2 = {\rm Str} \, s_I^2/R^2 = 0$ but   ${\rm Str} \, {\cal M}_0^4 = {\rm Str} \, s_I^4/R^4 \ne 0$.  
Then the leading contributions to the logarithmically divergent $V_{1,red}$ and to the finite $V_1$ are those of order $s_I^4$. 
Equating the two, the effective cutoff for the reduced theory is
$\Lambda = e^{25/3}/(16 \, \pi^4 \, R) \simeq 2.7/R$,
in agreement with the expectations.
Finally, we can consider the case in which both ${\rm Str} \, {\cal M}_0^2 = {\rm Str} \, s_I^2/R^2 = 0$ and  ${\rm Str} \, {\cal M}_0^4 = {\rm Str} \, s_I^4/R^4 = 0$.
In such a case, $V_{1,red}$ and $V_1$ are both finite, with    
$V_{1,red} 
=  {\rm Str} ( {\cal M}_0^4 \, \log {\cal M}_0^2 )/ (64 \, \pi^2)
=  {\rm Str} ( s_I^4 \, \log s_I^2 )/ (64 \, \pi^2 \, R^4)
$
and $V_1$ receiving contributions only from the last line of Eq.~(\ref{V1exp}), therefore coinciding with $V_{1,red}$ up to corrections of order $s_I^6$ or higher.   

In the $N=2,4,8$ cases,  $V_1$ in (\ref{V1res}) is always negative semidefinite, vanishes only for unbroken supersymmetry and for fixed $R$ is minimised by half-integer twists. 
At any individual KK level $n \ne 0$, the supertraces are identical to those in the reduced theory ($n=0$). 
In $N=2$, where there is only a single twist $a$,  ${\rm Str} \, {\cal M}_0^2 =  {\rm Str} \, {\cal M}_n^2  =  {\rm Str} \, s_I^2 /R^2 = - 8 a^2/R^2<0$.
In $N=4$, where there are two independent twists $a_i$ $(i=1,2)$, ${\rm Str} \, {\cal M}_0^2 =  {\rm Str} \, {\cal M}_n^2 =  0$, as expected, and ${\rm Str} \, {\cal M}_0^4 =   {\rm Str} \, {\cal M}_n^4 =  {\rm Str} \, s_I^4/R^4 = 72 \, a_1^2 \, a_2^2/R^4>0$.
In $N=8$, as originally described in \cite{Cremmer:1979uq}, there are four independent twists $a_i$ $(i=1,2,3,4)$.
In the reduced theory, it is known that ${\rm Str} \, {\cal M}_0^{2k} = 0$ ($k=1,2,3$) and 
${\rm Str} \, {\cal M}_0^8 =  {\rm Str} \, s_I^8/R^8 = 40320 \, a_1^2 \, a_2^2 \, a_3^2 \, a_4^2 / R^8$.
We find that the same result holds true at any fixed KK level $n \ne 0$ in the compactified theory:
${\rm Str} \, {\cal M}_n^{2k} = 0$ ($k=1,2,3$) and 
${\rm Str} \, {\cal M}_n^8 =  {\rm Str} \, s_I^8/R^8 = 40320 \, a_1^2 \, a_2^2 \, a_3^2 \, a_4^2 / R^8$.
In the limit $|a_i| \ll 1$  (for all $i=1,2,3,4$), the contributions of the KK modes become negligible and $V_1 \simeq V_{1,red}$, up to  corrections of order $s_I^8$ and higher.
For half-integer twists, $V_1 \simeq - 0.0125/R^4$, not too far from $V_{1,red} \simeq - 0.0184/R^4$.

In the case of a single extra dimension compactified on a circle, considered in the above example, the Scherk--Schwarz twist leading to supersymmetry breaking corresponds to a global continuous R-symmetry of the higher-dimensional action. 
We can smoothly take the limit of small twist parameters, $s_{\alpha} \ll 1$, and continuously connect the full effective potential $V_1$ of the compactified theory with the effective potential $V_{1,red}$ of the reduced theory.
Moreover, the tree-level spectrum depends on a single modulus $R$. 
It is then not surprising that $V_1$ and $V_{1,red}$ have a very similar behaviour, and that in the $N=8$ case the disappointing result obtained for $V_{1,red}$ is now corrected but qualitatively confirmed also for $V_1$. 

However, in the case of twisted tori with three or more extra dimensions, already mentioned in section~\ref{sec:dimred}, the twist parameters are quantized and the reduced theory can either be a mere consistent truncation or a genuine effective theory \cite{DallAgata:2005zlf, Grana:2013ila}, depending on whether the compactification manifold is homogeneous or not. 
In the first case, the three-dimensional manifold is still a torus and supersymmetry appears to be broken in the reduced truncated theory but remains unbroken in the full compactified theory. 
In such a case, we expect $V_1=0$, i.e. perturbative stability for the Minkowski vacuum of the full higher-dimensional theory, which is indeed supersymmetric, although $V_{1,red}$ could have a completely different qualitative behaviour.   
Also in the second case, the tree-level field-dependent spectrum of the compactified theory may have a non-trivial structure, with masses depending on several moduli and $V_1$ could exhibit significantly different qualitative features from $V_{1,red}$.
 
The simplest scenario we can consider is three extra dimensions, with the internal manifold given by a twisted torus corresponding to the freely acting orbifold $T^3/Z_k$ ($k=2,3,4,6$), where $T^3 \sim T^2 \times S^1$ and a discrete translation along the circle $S^1$ is combined with a discrete rotation of $T^2$.
In the $N=8$ case, this can be studied both at the field theory \cite{GDAFZ2025} and at the string theory \cite{Acharya:2020hsc} level, although in the string theory case one needs to take $M_C \ll M_S$ to avoid tachyonic winding states.
Both descriptions, in the cases studied so far, confirm the one-loop instability of the flat directions, with a negative $V_1$ scaling as an inverse power of the $S^1$ radius $R$. 
The above studies can be extended to supersymmetry-breaking compactifications on $n$-dimensional ($3 \le n \le 7$) freely acting orbifolds, for which a systematic classification is still missing.  
All these flat compactifications of higher-dimensional supergravities do not require D-branes and/or orientifolds to comply with known no-go theorems for Minkowski vacua of higher dimensional theories \cite{deWit:1986mwo,Maldacena:2000mw}: they are therefore the obvious next step to explore, and probably the last one where stringy effects can be consistently neglected.

\chapter*{Acknowledgments}
\vspace*{-3.0cm}
I thank Anna Ceresole and Gianguido Dall'Agata for their kind invitation to contribute to this volume, and Costas Bachas for useful comments.
I also thank
A.~Brignole,
F.~Catino, 
G.~Dall'Agata,
J.P.~Derendinger,
S.~Ferrara,
F.~Feruglio,
G.~Inverso,
C.~Kounnas$^\dagger$,
H.~Luo,
I.~Pavel,
M.~Petropoulos,
M.~Porrati
and
G.~Villadoro
for enjoyable collaborations on no-scale supergravity.

\end{document}